\documentclass[11pt]{article}
\usepackage{graphicx}
\usepackage{epsfig}
\usepackage{amsfonts}
 \usepackage{amssymb}
\usepackage[dvips]{color}
\date{}

\newcommand{\insertplot}[5]{\begin{figure}
 \hfill\hbox to 0.05in{\vbox to #5in{\vfill
 \inputplot{#1}{#4}{#5}}\hfill}
 \hfill\vspace{-.1in}
 \caption{#2}\label{#3}
 \end{figure}}
 \newcommand{\inputplot}[3]{
 \special{ps: plotfile #1}
}
\newcounter{fig}

\newcommand{\ee}{\end{equation}}
\newcommand{\eea}{\end{eqnarray}}
\newcommand{\be}{\begin{equation}}
\newcommand{\bea}{\begin{eqnarray}}

\begin{document}
\title{
Scalarized dyonic black holes  in vector-tensor Horndeski gravity.
}
\author{
{\large Y. Brihaye}$^{\dagger}$ \textit{and}
{\large Y. Verbin} $^{\ddagger}$
\\
\\
$^{\dagger}${\small Physique-Math\'ematique, Universit\'e de
Mons, Mons, Belgium}
\\
$^{\ddagger}${\small Department of Natural Sciences, The Open University of Israel, Raanana, Israel}
}
\maketitle
\begin{abstract}
The  vector-tensor Horndeski theory is  supplemented by a real, massless scalar field non-minimally coupled to the Horndeski interaction term.
The generic dyonic Reissner-Nordstrom solutions characterized by electric and magnetic charges, acquire an instability and bifurcate
into electromagnetically charged solutions  presenting an horizon and a cloud of scalar hairs at specific values of the non-minimal coupling constant.
These hairy black holes  are studied in terms of the physical parameters of the model and  the critical phenomena limiting
 their domain of existence are discussed.
\end{abstract}


\section{Introduction}
\setcounter{equation}{0}
In the effort aimed to understand the dark matter and dark energy problem in the Universe, numerous modifications -or extensions-
of General Relativity (GR) have been explored in the last decades.
One of the most natural strategies of extension consists of two steps~:
(i) supplementing fields of GR  by one (or more) fields inspired by elementary particle physics:
scalar, vector or spinor fields;
(ii) adding non minimal interactions between the gravitational fields and the extra matter fields; these new interactions
beeing charaterized by one or more coupling constants.
\\
The elaboration of step (ii) is limited by  several essential principles like Lorentz-covariance and second order field equations.
In this respect the early work of G. Horndeski turns out to be crucial. In Refs. \cite{Horndeski:1974wa}, \cite{Horndeski:1976gi} respectively
the minimal Einstein-Hilbert action of GR was extended by a scalar field and by a vector field  with the request that the field equations remain
covariant and of second order in the derivatives with respect to space-time coordinates.
Although the scalar-tensor theory elaborated in \cite{Horndeski:1974wa} contains a lot of arbitrariness, the vector-tensor lagrangian of \cite{Horndeski:1976gi}
is relatively simpler, namely it involves a single new interaction term with a corresponding coupling constant.
In particular, the model obtained in \cite{Horndeski:1976gi} encompasses the Einstein-Hilbert-Maxwell lagrangian;
the study of spherically symmetric solutions of the corresponding equations were addressed in \cite{muller} and revisited
recently in \cite{Verbin:2020fzk}.

These various generalizations of GR offer multiple directions of research; the most exciting being the study of the
effects of the new interaction on the
large structures of the Universe. Parallelly, they provide interesting grounds to  circumvent the No-hair and/or uniqueness theorems
\cite{no_hole_old}, \cite{no_hole_new}
which limit the possible black holes solutions in minimal GR and in its extension by the standard Maxwell term.

Families of new black holes were shown to exist in several models of scalar-tensor gravity,
see e.g. \cite{Herdeiro:2014goa},\cite{Sotiriou:2013qea},\cite{Sotiriou:2014pfa}.
These solutions present a horizon
surrounded by a cloud -a shell- of scalar field vanishing far away from the horizon region.  More information and references
 can be found  in \cite{Sotiriou:2015pka},\cite{Herdeiro:2015waa},\cite{Volkov:2016ehx} reviewing the topic.

A specific subclass of the general scalar-tensor Horndeski lagrangian was
studied in \cite{Doneva:2017bvd}, \cite{Silva:2017uqg}, \cite{Antoniou:2017acq}. For these models it was demonstrated that
the hairy configurations appear when the coupling constant, say $\alpha$,
takes values in a specific interval  and that the scalar hairs disappear
at the approach of some critical value, say $\alpha_c$. For the other values of $\alpha$ the
field equations are incompatible with a non-trivial scalar field and admit exclusively the standard black holes (Schwarzschild, Reissner-Nordstrom, Kerr).
Hairy black holes appearing in this way are said to scalarize spontaneously.
The idea of spontaneous scalarization was extended with different kinds of non-minimal interacting terms
in \cite{Herdeiro:2018wub},\cite{Fernandes:2019rez} and \cite{Brihaye:2019kvj}. In these  papers, the scalar field is non minimally coupled
ether to space-time  via  geometric invariants (e.g. the Ricci or Gauss-Bonnet term) or to electromagnetism via the Maxwell term.

The purpose of this present paper is to emphasize the existence of spontaneous scalarization in the Vector-Tensor model of \cite{Horndeski:1976gi}
suitably extended by a scalar field. In a sense our model is inspired by \cite{Fernandes:2019rez}
but here the scalar field is non-minimally coupled to the 'Horndeski-interaction' which mixes both gravity and electromagnetism.
Hairy black holes in such a model were constructed in \cite{Brihaye:2020oxh} for a purely electrostatic vector potential. In this paper we extend our previous work for dyonic electromagnetic field, i.e. presenting both electric and magnetic charges.

The paper is organized as follows. The model, field equations and ansatz and various general properties are specified in Sec. 2.
Two specific limits of the equations (the probe and the linearized scalar limits)
are studied respectively in Sec. 3 and Sec. 4.
 It is shown how the critical value of the coupling constant, say $\alpha_c$, where spontaneous scalarization
occurs can be determined via the linearized equations. A special emphasis is set on the dependance of $\alpha_c$
on the electric and magnetic charge
of the underlying dyonic  Reissner-Nordstrom background.
Sec. 5 contains our results for the hairy black hole solutions (``HBH'' for short) of the full system of equations with an emphasis on the effects of the magnetic charge
on the solutions of \cite{Brihaye:2020oxh}.

\section{Model and classical equations }
\setcounter{equation}{0}
We are interested in classical solutions associated with
Einstein-Maxwell-Klein-Gordon lagrangian extended by a non-minimal
coupling between the gravitational, vector and scalar fields inspired by the work of G. Horndeski.
The action considered is  the form
\be
   S = \int d^4 x \sqrt{-g} \bigg[ \frac{1}{2\kappa} R - \frac{1}{2} \nabla_{\mu} \phi \nabla^{\mu} \phi
	- \frac{1}{4} F_{\mu \nu}F^{\mu \nu} + H(\phi) {\cal I}(g,A)  \bigg]
\label{lagrangian}
\ee
where $R$ is the Ricci scalar,
 $F_{\mu \nu}$ in the electromagnetic field strength
 and $\phi$ represents a real scalar field.
The  tensor-vector gravity theory  is  characterized by
non-minimal coupling of the vector field to the geometry ${\cal I}(g,A)$.
\be
\label{action}
{\cal I}(g,A) = -\frac{1}{4} (F_{\mu \nu} F^{\kappa \lambda} R^{\mu \nu}_{\phantom{\mu \nu} \kappa \lambda}
                          - 4 F_{\mu \kappa} F^{\nu \kappa} R^{\mu}_{{\phantom \mu} \nu}
													+ F_{\mu \nu} F^{\mu \nu} R ) \ .
\ee
Following the spirit of many recent works, the  tensor-vector gravity lagrangian has been augmented
by a real scalar field $\phi$ and
this  scalar is coupled  to the Horndeski interaction term via the coupling function $H(\phi)$ that we choose according to
\be
\label{coupling}
               H(\phi) =  \alpha \phi^2
\ee
where $\alpha$ represents the non-minimal coupling constant which in most cases we will take to be positive.

Recently, several hairy black holes have been constructed  in models of the type (\ref{action}) above,
mainly by choosing for ${\cal I}(g)$ the Gauss-Bonnet invariant or the Faraday-Maxwell term  ${\cal I}(A) = F^2$.
In these works  several forms of the function $H(\phi)$ were  suggested in order to construct neutral hairy black holes,
see e.g. \cite{Doneva:2017bvd}, \cite{Silva:2017uqg},  \cite{Antoniou:2017acq}, \cite{Antoniou:2017hxj}
as well as charged ones \cite{Brihaye:2019kvj}.

\subsection{Ansatz and field equations}
We will be interested in spherically symmetric solutions and adopt  a  metric of the form
\be
     ds^2 = - f(r) a^2(r) dt^2 + \frac{1}{f(r)} dr^2 + r^2 d \Omega_2^2
\ee
completed by the scalar field    and a vector field respectively of the form
\be
\phi(x^{\mu}) = \phi(r) \ \ , \ \ A_{\mu} dx^{\mu}=V(r) dt  + P(1-\cos(\theta)) d \varphi
\ee
where the  constant magnetic charge $P$  can be assumed to be positive without loss of generality.

Substituting the ansatz in the field equations, the  system can be reduced to a set of four  non linear differential
equations (plus a constraint) in the functions $f(r), a(r), V(r)$ and $\phi(r)$.
These equations  can be obtained directly by varying the following effective lagrangian
\bea
\label{lag_eff}
{\cal L}_{eff} &=& \frac{a(1-rf'-f)}{\kappa} - \frac{afr^2}{2} \phi '\;^2 + \frac{r^2}{2a}(V')^2 -\frac{P^2 a}{2 r^2} \nonumber \\
               &+& \alpha \Biggl(
							\phi^2 \frac{1-f}{a}(V')^2 +  P^2 \frac{af'+2 a'f}{r^3}(\phi^2 - r \phi\phi')
							             \Biggr)
\eea
The equation for $V(r)$ can be integrated easily to give
\be
V' = \frac{ Q a}{ r^2 + 2 \alpha  (1-f) \phi^2}
\label{FEqCharge}
\ee
where $Q$ is the integration constant which is proportional to the charge of the solution. The field equations for the metric functions and for the scalar field  take the following form after using Eq. (\ref{FEqCharge}):
\bea \label{FEqMetricf}
1-f-r f'  -\frac{\kappa }{2} r^2 \left(\frac{Q^2}{\left(r^2+ 2 \alpha  (1-f) \phi^2\right)^2}+\frac{P^2}{r^4}+
f \phi '\;^2\right) + \hspace{5.6cm}\\
   \alpha  \kappa  \phi ^2 \left(\frac{P^2 \left(6 f-r f'\right)}{r^4}-\frac{ Q^2(1-f)}{\left(r^2+ 2 \alpha  (1-f) \phi^2\right)^2}\right)+
\frac{2 \alpha  \kappa  P^2 }{r^2}\left(\frac{\phi  \phi ' \left(r f'-8 f\right)}{2 r}+f
   \left(\phi  \phi ''+\phi '\;^2\right)\right)=0 \nonumber
\eea
\bea
\frac{ r a'}{a}-\frac{\kappa }{2} r^2  \phi '\;^2+ \alpha  \kappa  \phi ^2 \left(\frac{P^2}{r^4} \left(\frac{r a'}{a} +3 \right)-\frac{ Q^2}{\left(r^2+ 2 \alpha  (1-f) \phi^2\right)^2}\right)+ \nonumber \\
   \frac{\alpha  \kappa  P^2 }{r^2}\left(\phi '\;^2  -\frac{\phi  \phi ' }{r}\left(\frac{r a'}{a}+4 \right)+\phi  \phi ''\right)=0
   \label{FEqMetrica}
\eea
\bea
\left(1+\frac{\alpha  \kappa  P^2 \phi ^2}{r^4}\right) \left(f \phi ''+\left(f \left(\frac{r a'}{a}+2\right)+r f'\right)\frac{\phi ' }{r}\right)+ \frac{4 \alpha ^2  \kappa  P^2 Q^2 f \phi ^2  \left(r^2-2 \alpha  (1-f) \phi^2\right)}{r^5 \left(r^2+ 2 \alpha  (1-f) \phi^2\right)^3} \phi ' - \nonumber \\
 \frac{2 \alpha  P^2  \left(f a'+a f'\right)\phi}{r^5 a } + \frac{\alpha  Q^2  \left(\frac{\kappa  P^2 }{r^4}\left(r^2+ 2 \alpha  \phi ^2 \left(1-f+\frac{ r a'}{a}f \right)\right)+2 (1-f)\right)\phi }{r^2 \left(r^2+ 2 \alpha  (1-f) \phi^2\right)^2} +  \nonumber \\
  \frac{\alpha  \kappa  P^4 \phi }{r^8} -    \frac{\alpha  \kappa  P^2  f \phi  \left(\phi
   '\right)^2}{r^4}+\frac{2 \alpha ^2 \kappa  P^2 Q^2 \phi ^3 \left(f' \left(r^2+2 \alpha  (1+f) \phi ^2\right)-4r f \right)}{r^5 \left(r^2+ 2 \alpha  (1-f) \phi^2\right)^3} =0 \hspace{1cm}
\label{FEqScalar}
\eea

The three final equations are of the first order for the functions $f(r)$, $a(r)$ and of the second order for $\phi(r)$;
they depend on the coupling constant $\alpha$ and of the two charge parameters $Q,P$.

\subsection{Boundary conditions}
For a fixed choice of $Q,P$ and $\alpha$ four boundary conditions need to be fixed to specify a solution.
Because we  look for localized, asymptotically flat solutions, we require
\be
\label{cond_infty}
     f(r \to \infty) = 1 \ \ \ \ , \ \ \ \ a(r \to \infty) = 1 \ \ \ \ , \ \ \ \ \phi(r \to \infty) = 0 \ \ .
\ee
Imposing a regular horizon at $r=r_h$ needs $f(r_h)=0$.
The equation of the scalar field is then singular in the limit $r\to r_h$; obtaining regularity requires a very specific relation
between the values $\phi(r_h)$ and $\phi'(r_h)$. Examining the equations, the conditions for a regular black hole at $r_h$ turn out to be~:
\be
\label{cond_regular}
   f(r_h) = 0 \ \ , \ \   \phi'(r_h) = \frac{2\alpha \phi_h {\cal A}}{(1+2\alpha \phi_h^2){\cal B}} \ \ ,
\ee
where ${\cal A},{\cal B}$ are involved polynomials in $Q,P$ and $\phi_h \equiv \phi(r_h)$~:
\bea
     {\cal A} &=& 2(P^2 Q^2 + Q^2 + P^4-P^2) + \alpha \phi_h^2 (-Q^4 P^2 + 12 P^2 Q^2 + 4 Q^2 + P^6 + 12 P^4 - 12 P^2) \nonumber \\
		   &+& 2 \alpha^2 P^2 \phi_h^4(P^2 Q^2 + 8 Q^2 + 3 P^4 +12 P^2 - 12)  \nonumber \\
			 &+&  4 \alpha^3 \phi_h^6 P^2(P^2 Q^2 + 3 P^4 + 5 P^4 - 4 ) + 8\alpha^4 \phi_h^8 P^6 \ \ , \label{expression for A} \\
     {\cal B} &=& P^2 + Q^2 - 2 + \alpha \phi_h^2(P^2 Q^2 + 2 Q^2 + P^4 +2 P^2 - 8) \nonumber \\
		   &+& 2 \alpha^2 \phi_h^2 ( P^2(P^2 Q^2 + 2 Q^2 + P^2) + \phi_h^2(P^2 Q^2 + 2 P^4 - 2 P^2 - 4)) \nonumber \\
			 &+& 4 \alpha^3 P^2 \phi_h^4 (P^2 Q^2 + 2 P^4 + \phi_h^2 (P^2-2)) + 8 \alpha^4 \phi_h^6 P^6 \ \ ,
\label{expression for B} \eea
where for compactness we set $\kappa=1,r_h=1$ by appropriate rescalings.
The boundary value problem is then fully specified by two of the three conditions of (\ref{cond_infty}) and by (\ref{cond_regular}).
Enforcing a hairy black hole will imply in addition
setting a value $\phi(r_h) \neq 0$ for the scalar field.
Henceforth the number of conditions to be imposed on the boundary needs to exceed four (the number allowed by the order of the equations)
the generic solutions will therefore exist only via a specific  relation between the parameters $\alpha, Q, P$
and $\phi(r_h)$.
The regularity condition  (\ref{cond_regular}) further implies ${\cal B}(\alpha, P, Q, \phi_h ) \neq 0$; it will be pointed out in the next section that
the hypersurface ${\cal B}=0$ plays an important role in the domain of existence of the solutions. In addition,
$f'(r_h)$ will provide also some insight and we note that it is given by:
\be
\label{expression for f'}
   f'(r_h) = -\frac{{\cal B}}{2{\cal C}},
\ee
where
\bea
{\cal C}=1 + 2\alpha\phi_h^2 (P^2+2) - 2\alpha^2\phi_h^2 P^4  +
\alpha^2 \phi_h^4 ( P^4+ 8 P^2+4) +  \hspace{1cm} \nonumber \\
2 \alpha^3\phi_h^4 P^4 (Q^2 - 4) +
  4\alpha^3\phi_h^6 P^2 ( P^2+2) -   8 \alpha^4\phi_h^6 P^4 + 4 \alpha^4\phi_h^8 P^4 .
\eea


\subsection{Asymptotic form}
These hairy black holes   can  be characterized by their mass $M$, their electromagnetic  charges $Q,P$ and the scalar charge $D$.
They are related respectively to the asymptotic decay
of the functions $m(r)=r(1-f(r))/2$, $V(r)$ and $\phi(r)$. The asymptotic behavior of the various fields reads~:
\bea\label{Asymptotics1-HRN-Scalar} \nonumber
      m(r) &=& M - \frac{\kappa (Q^2+P^2+D^2)}{4 r} - \frac{\kappa M D^2}{4 r^2} +   O(1/r^3) \\
      a(r) &=& 1 - \frac{ \kappa D^2}{4 r^2}   - \frac{2\kappa M D^2}{3 r^3}   +  O(1/r^4)
\eea
\bea\label{Asymptotics2-HRN-Scalar}
     V(r)&=&  - \frac{Q}{r} + \frac{\kappa Q D^2}{12r^3} + \frac{\kappa M Q D^2}{6r^4}+ \dots  \\ \nonumber
     \phi(r) &=& \frac{D}{r} +\frac{M D}{r^2} + \frac{D( 16 M^2-  \kappa(2 P^2 + 2 Q^2 + D^2 )}{12 r^3}   +  O(1/r^4)   \ \ .
\eea
The  temperature $T_H = a(r_h) f'(r_h)/(4 \pi)$ further characterizes the solutions. Note that the scalar charge $D$ is an independent charge and is not fixed by $M$, $P$ and $Q$. The non-minimal coupling constant appears explicitly only in the higher order terms of the asymptotic expansion, but its effect is of course evident from non-vanishing scalar charge.

\section{Scalar clouds around the Schwarzschild black hole }
\setcounter{equation}{0}
The full system of field equations is highly non-linear and
quite involved. We solved it by using the numerical routine COLSYS \cite{COLSYS} (more details are given at the beginning of Sec. 5). However, some insight about the pattern of solutions can be gained through the study of some special limits.
A particularly simple -but instructive- case consists of considering the equations in the  probe limit, i.e. setting $\kappa=0$ in the field equations (\ref{FEqCharge})--(\ref{FEqScalar}). This limit describes the situation where the ``back reaction'' of the scalar and gauge fields is negligible and is not taken into account.
The Einstein equations are solved, leading naturally to the space-time of a  Schwarzschild black hole~: $a(r)=1 \ , \ f(r) = 1 - r_h/r$;
leaving  a system of two equations for the vector and scalar fields~:
\be
\label{cloud}
   (r^2 + 2 \alpha (1-f) \phi^2)V' = Q \ , \ \ -(r^2 f \phi')' = \alpha \phi (W_e + W_m) \ , \ \ W_e = 2(1-f)(V')^2 \ , \ \ W_m = \frac{P^2 f''}{r^2} \ ,
\ee
where the vector field equation is integrated once  already in terms of the integration constant $Q$, leaving the single  non-trivial equation for $\phi$.
This non-linear equation can be solved numerically with the appropriate boundary conditions, for instance~:
\be
       \phi'(r_h) = 2\alpha \phi(r_h) \left( \frac{P^2}{r_h^5} - \frac{Q^2}{r_h(r_h^2+2\alpha \phi(r_h)^2)^2}\right) \ \ , \ \ \ \phi(r\to \infty) = \frac{D}{r} \ .
\ee
The solutions correspond to  clouds of scalar and electromagnetic fields in the background of the Schwarzschild black hole.

Prior to presenting these cloud solutions, it is instructive  to determine the region in the parameter space $\alpha,Q,P$
where the scalar clouds will emerge as bifurcations from the purely Schwarzschild space-time.
For this aim,   it is sufficient to consider an infinitesimal scalar function, i.e. $\phi(r) \ll 1$ and to solve the
linearized version of Eq. (\ref{cloud})~:
\be
     -(r^2 f \phi')' = 2\frac{\Omega}{r^5} \phi \ \ , \ \ \Omega \equiv  \alpha r_h( Q^2 - P^2) \ ;
\ee
which becomes the single relevant equation.
In spite of some efforts, we were unable to bring this simple equation to a form allowing analytical solutions
 but it is easy to show (numerically) that it admits a nodeless solution regular at the horizon
for $\Omega \equiv \Omega_0 \approx 1.2514$. This result indicates that
 clouds of scalar and electromagnetic fields are expected to bifurcate from the Schwarzschild black hole when the non-minimal coupling and charges become related by this condition. Later, the coupling to gravity will return by relaxing the probe limit (i.e. revert to $\kappa>0$) and these solutions will deform into the hairy black holes discussed in Sec. 5.

\begin{figure}[t!]
\begin{center}
{\includegraphics[width=10cm, angle=-00]{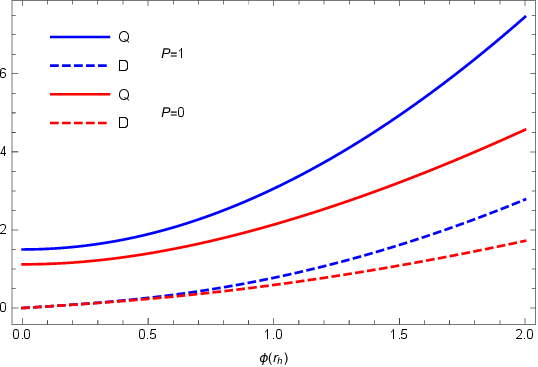}}
\caption{{\small The dependance of the charges $Q,D$ for scalar clouds on $\phi(r_h)$ for $\alpha=1$ and for two values of the magnetic charge $P$.
\label{gamma_1}
}}
\end{center}
\end{figure}

We can show numerically that scalar clouds indeed exist as solutions to Eq.(\ref{cloud}). They exist on a geometrical background of a Schwarzschild black hole with an additional radial electric field.
For simplicity, we limit the construction to the case where the scalar
field presents no nodes  and set $r_h=1$ by a suitable rescaling.
We concentrate in the case $\alpha >0$, because  for $\alpha < 0$ the  electric field may present a singularity  (see Eqs. (\ref{FEqCharge}) and (\ref{cloud})).
Increasing progressively the value $\phi(r_h)$, families of configurations with non trivial scalar and electromagnetic field can be constructed
in the domain  $Q,P$ such that $\alpha(Q^2-P^2) > \Omega_0 \approx 1.2514$.
The value $\phi(r_h)$ is uniquely fixed by the choice $\alpha, Q,P$. Our results also suggest  that the family of solutions corresponding to
fixed values of $\alpha,P$ exist for arbitrary values of the electric charge $Q$.
The electric and scalar charges $Q,D$ respectively  increase monotonically with $\phi(r_h)$ as demonstrated by Fig. \ref{gamma_1};
the limiting values of $Q$ for given $P$ in the limit  $\phi(r_h) \to 0$ is read off from the relation $\Omega = \Omega_0$.

\section{Solutions in a Reissner-Nordstrom black hole background }
\setcounter{equation}{0}

The  next to simplest case prior to addressing the full system of field equations consists of considering the limit
of an infinitesimally small scalar field, or formally, we restore the gravitational constant $\kappa$ and
 consider the equations  linearized in the scalar field. In this way the
Einstein-Maxwell equations become independent on the scalar field and admit  the family of dyonic  Reissner-Nordstrom (RN) black holes as generic solutions~:
\be
     f(r) = 1 - \frac{2M}{r} + \frac{\kappa(Q^2 + P^2)}{2r^2} \ \ , \ \ \ a(r) = 1 \ \ , \ \ V(r) =  - \frac{Q}{r}
\ee
characterized by the mass parameter $M$   and the electric and magnetic charges $Q,P$.
The event horizon radius $r_h$ is then defined as the largest root of the function $f(r)$, i.e. $f(r_h)=0$ leading to $M = (2r_h^2+\kappa(Q^2 + P^2))/(4 r_h)$.
The horizon is simple  for $\kappa(Q^2 + P^2) < 2 r_h^2$ and  becomes double for $\kappa(Q^2 + P^2) = 2 r_h^2$ which defines the extremal RN black hole (ERN).
In the numerical construction of the solutions, the scale of the radial variable was set it such a way that $\kappa = 1$ and $r_h=1$.

The relevant equation for the scalar field in this limit reduces to the  linear equation
\be
\label{probe_eq}
   - (r^2 f  \phi')' = \alpha W \phi \ \ , \ \ W(r) = 2(1-f)(V')^2 + \frac{P^2}{r^2} f'' \ \ .
\ee
It needs to be solved with a condition which guarantees the regularity of the field $\phi(r)$ at $r=r_h$ and the condition that the scalar field vanishes at infinity;
     because the equation is linear the normalization of the function $\phi$ has to be fixed as well.
This  leaves us to solve the second order equation Eq. (\ref{probe_eq}) with three boundary conditions
\be
 \phi(r_h)= 1 \ \ , \ \
\frac{\phi'}{\phi}(r_h) = 4\alpha \frac{\kappa P^2(Q^2 + P^2)+r_h^2(Q^2 - P^2)}{r_h^5(\kappa(Q^2 + P^2)-2 r_h^2)} \ \ , \ \ \phi(r \to \infty) = \frac{D}{r}
\ee
Eq. (\ref{probe_eq}) can then be seen as a condition for the linear operator $d_r(r^2 f d_r) + \alpha W(r)$ to admit a bound state with null eigenvalue.
In order to fulfill  the three boundary conditions, the parameter $\alpha$ has to be fine tuned for a fixed choice of  $Q,P$,  or equivalently, for fixed $\alpha$ and one of the electromagnetic charges, the other one should be fine tuned. In other words, the
solutions will occur on a special surface of the three parameter space,  say
\be
\label{surface}
   {\cal S}(\alpha, P, Q) = 0
\ee
Integrating the LHS equation of Eq.(\ref{probe_eq}) on $[r_h, \infty]$ and using the relevant boundary conditions and asymptotic behavior of Eqs. (\ref{cond_infty}), (\ref{cond_regular}) and (\ref{Asymptotics2-HRN-Scalar}) leads to
\be
         D = \alpha \int_{r_h}^{\infty} W(r) \phi(r) \ dr \ .
\ee
This relation indicates that solutions with no-node (which then have  $D > 0$) exist provided the potential $\alpha W(r)$ is not too negative.
In the pure electric case (i.e. for $P=0$) the potential $W$ is positive definite
so that solutions only exist for $\alpha > 0$; for $P>0$ the potential is negative in some regions so that solutions can exist for both signs of $\alpha$.

\begin{figure}[b!]
\begin{center}
{\includegraphics[width=9.5cm, angle = -00]{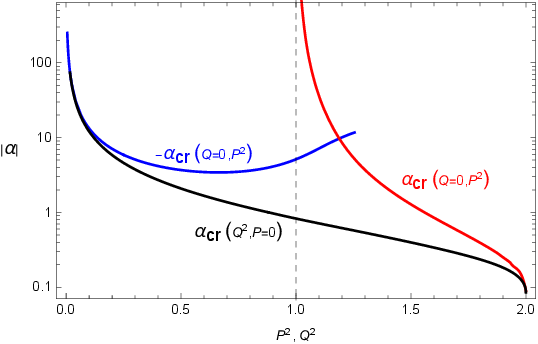}}\vspace{5mm}
{\includegraphics[width=9.5cm, angle = -00]{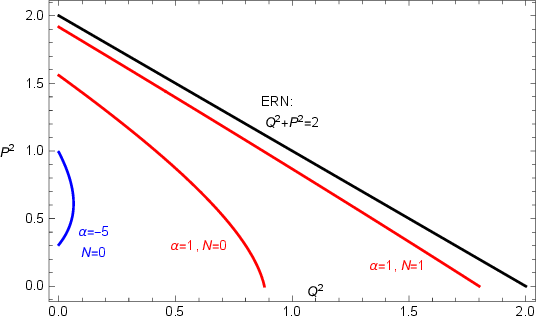}}
\end{center}
\caption{ {\small Up:
The critical value of $\alpha$ as function of $P^2$ (resp. $Q^2$) for RN solutions with $Q=0$ (resp. $P=0$).
 Bottom:  Curves in $Q^2,P^2$ plane with fixed $\alpha$. The critical lines of existence of HBH with zero ($N=0$) and one ($N=1$) node corresponding to $\alpha=1$ and  HBH with zero ($N=0$) nodes corresponding to $\alpha=-5 $. Also seen is the curve for the extremal RN solutions (ERN-black line).
\label{domain_p2_q2}
}}
\end{figure}

The possible values of $\alpha$ for no-node solutions are plotted on the upper part of Fig.\ref{domain_p2_q2} both for the purely electric case ($P=0$) and for the purely magnetic case ($Q=0$). Recall that the figure presents the rescaled parameters (i.e. in units where $r_h = \kappa = 1$). We notice the following behavior:
\begin{itemize}
\item In the pure electric case (black line) the solutions exist for $Q^2 \in ]0,2[$. As expected, the lower the charge $Q$  is, the higher should be
the constant $\alpha$ allowing for a bound state (more precisely: $\lim_{Q^2 \to 0} \alpha_c = \infty$); at the approach of the extremal RN solution  we
found $\lim_{Q^2 \to 2} \alpha_c \approx 0.1$.
Hairy black holes with purely electric charge are therefore expected to exist for $\alpha \geq 0.1$.
\item In the pure magnetic case ($Q=0$) the potential reduces to $W(r)=P^2(3P^2-(2+P^2)r)/r^6$.
This function is negative for $P<1$, accordingly only solutions
with $\alpha < 0$ exist in this case; these are represented by the blue line on the figure \ref{domain_p2_q2} (upper part).
For  $P^2 \in ]1,2[$ the potential is positive close to the horizon and negative asymptotically;
solutions  were found for both signs of $\alpha$, as seen by the blue and red
lines on the upper panel. Notice~: the blue line stops because the derivative $\phi'(r_h)$ diverges when $P$ becomes too large.
\end{itemize}

On the lower part of Fig. \ref{domain_p2_q2} the relation between $P$ and $Q$ is
depicted for $\alpha = 1$ for the solutions with  zero and one node (red lines labeled
$N=0$ and $N=1$ respectively) and for solutions corresponding to $\alpha = -5$ (blue line, only for no node solutions in this case).
The RN black holes exist below the black line $Q^2+P^2=2$ and become extremal on this line.
The figure strongly suggests that  hairy black hole solutions will bifurcate from RN black holes at the points corresponding to the red and blue curves.
This will be discussed in the next section.

\section{Hairy black holes }
\setcounter{equation}{0}
\begin{figure}[b!]
\begin{center}
{\includegraphics[width=8cm, angle = -00]{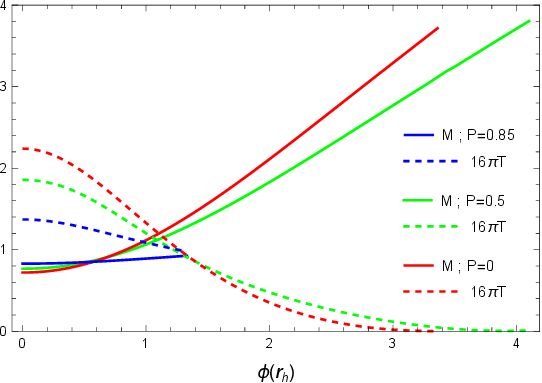}}
{\includegraphics[width=8cm, angle = -00]{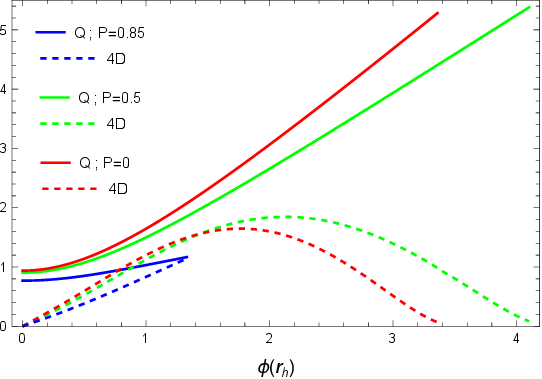}}
\end{center}
\caption{ {\small  The dependence on the horizon value of the scalar field of  mass, electric charge, scalar charge and temperature of the HBHs corresponding to $\alpha = 1$ with $P=0$, $P=0.5$ and $P=0.85$. Left: mass and temperature, Right: electric and scalar charges.
\label{QMDTvsPhiH}
}}
\end{figure}

We now discuss the  solutions obtained by solving the full system of  field equations (\ref{FEqMetricf})-- (\ref{FEqScalar}). 
The non-linear equations are solved numerically by using both the ODE solver of Mathematica and independently the Fortran solver COLSYS \cite{COLSYS}. This later routine is based on the Newton method of quasi linearization. At each step of the iteration a linearized problem is solved by using  a spline collocation at Gaussian points. The equations are solved for a particular grid
$x_0<x_1 < \dots x_j \dots x_F $
where $x_0, x_F$ are the boundary conditions and the mesh $x_j$ is fixed and adapted by the program.
The number of intermediate points is  typically $500$
and the last point $x_F$ set in sufficiently large
to capture the asymptotic decay of the fields. We typically imposed error tolerance in the range
$10^{-4} - 10^{-6}$  but the absolute errors of the collocation solutions turned out to be better, i.e. of the order $10^{-8}$.
For all cases the agreement between the two different numerical approaches was excellent.

By an appropriate choice of the scale of the matter fields (scalar and electromagnetic)  and of the radial variable, we can set $\kappa=1$ and $r_h=1$. For definiteness, we fixed  $\alpha=1$ which captures most of the qualitative features of the generic hairy black holes with $\alpha>0$. Indeed, scalarization appears in this system also for $\alpha<0$ as the discussion of Sec. 4 implies, but we choose to concentrate in the $\alpha>0$ as we did already in the purely electric case \cite{Brihaye:2020oxh}. However, we will give some attention to $\alpha<0$ solutions at the end of this section.

\begin{figure}[t!]
\begin{center}
{\includegraphics[width=9cm, angle=-00]{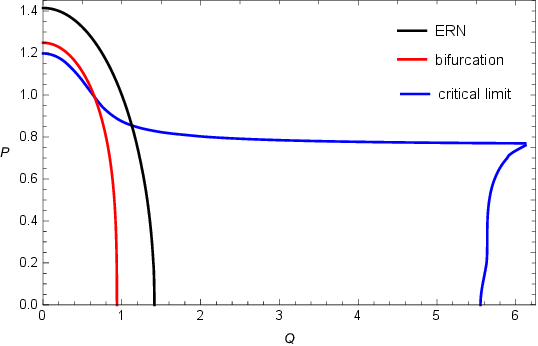}}
\caption{{\small Domain of existence in the $Q,P$ plane~: the HBHs exist between the bifurcation line (red) and the boundary line (blue). The RN black holes exist below the ERN black line. Notice that the ERN line looks elliptical because of different horizontal and vertical scales. It is the same line as the black line in Fig.\ref{domain_p2_q2}. Also the red line here is the same as the red bifurcation line ($\alpha=1$, $N=0$) of that figure.
\label{domain_2}
}}
\end{center}
\end{figure}

One possible strategy to construct the HBHs is the following:
(i) start from a point on the surface ${\cal S}$ defined in (\ref{surface}) with an infinitesimal value $\phi(r_h)$,
(ii) with two fixed values out of the triplet $\alpha, Q,P$ (e.g. $\alpha, P$),
increase progressively the value $\phi(r_h)$ (used as an initial condition) and compute
the corresponding value of the unfixed parameter ($Q$ in our example).
In this way, families of solutions parametrized by $\phi(r_h)$  and characterized by $\alpha, P, Q$ are obtained.
Each  solution can be further characterized by
the mass $M$, the  temperature $T_H$ and the scalar charge $D$ (we constructed only solutions with $\phi(r)$ presenting no node, node-solutions
likely exist as well but were not attempted to be constructed). Due to the occurrence of four parameters to vary,
the characterization of the full domain of existence in parameter space is quite involved.  Several features of the domain will be shown
on Fig. \ref{domain_2}, But before that, we present in Fig. \ref{QMDTvsPhiH} the behavior of the HBH various charges and temperature with varying $\phi(r_h)$ and their interrelations. First we notice the realization of the phenomenon of scalarization by the fact that the HBHs start with finite (non-zero) charge at $\phi(r_h)=0$. Further we notice that HBHs exist in a finite domain of $\phi(r_h)$ and of  electric and magnetic charges. Increasing the magnetic  charge first increases and then decreases the interval of $\phi(r_h)$ and of the electric charges  where HBNs exist. Increasing the electric  charge only decreases the interval of $P$  where HBNs exist. Finally we comment that we do not present the analogous plots to those of Fig. \ref{QMDTvsPhiH} of the $\phi(r_h)$-dependence for fixed values of $Q$ since it is less instructive: In most of the existence domain, lines of fixed $Q$ do not include the bifurcation from the RN solutions, but rather a family of HBHs which start from purely electric HBH (at $P=0$) and ends in a maximal magnetic charge.

Fig. \ref{domain_2} presents the domain where HBHs exist in the $Q,P$ plane where we fixed  $\alpha=1$. First, the HBHs bifurcate from the RN solutions on the red line of Fig. \ref{domain_2}; it corresponds to the case $\phi(r) \ll 1$ discussed in the previous section.
Approaching this line, the scalar field $\phi(r)$ tends to the null function, in particular
$\phi(r_h) \to 0$. The set of values of $Q,P$ allowing for HBHs can be explored in different ways but the basic behavior consists of increasing $\phi(r_h)$ with fixed $P$, then determining $Q$ numerically, or vice versa: fixing $Q$ and determining $P$.
These two ways to explore the set of solutions reveal that  different critical phenomenon constitute the limits of the domain of existence. The blue line 
in Fig. \ref{domain_2} represents the critical limit of the domain in the $Q,P$ plane;
it consists of two distinct parts corresponding to two different kinds of critical solutions.

\begin{figure}[h!!!]
\begin{center}
{\includegraphics[width=9cm, angle=-00]{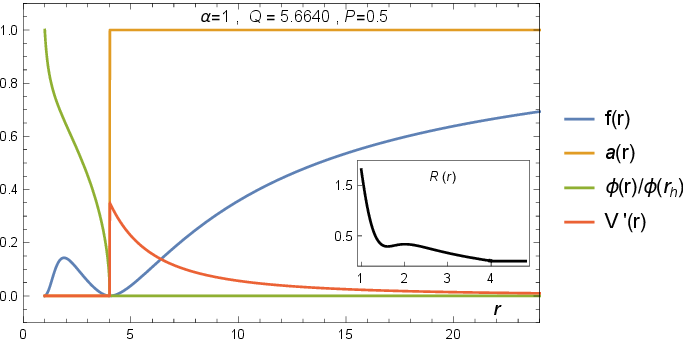}}
{\includegraphics[width=7.1cm, angle=-00]{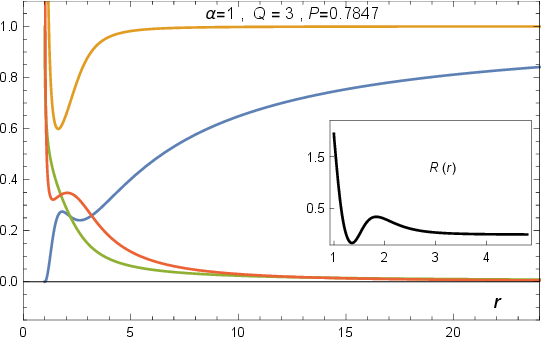}}
\caption{{\small The profiles of $f,a, \phi, V',$ and the Ricci scalar $R$ for the two kinds of extremal solutions corresponding to: Left: $P=0.5, Q_c= 5.6640$. Right: $Q=3, P_c= 0.7847$.
\label{profile_p_05} }}
\end{center}
\end{figure}
\begin{itemize}
\item For fixed $P$ such that $0 < P \lesssim 0.8$ and increasing $\phi(r_h)$ (or alternatively $Q$) the solutions stop at a maximal value $Q_c$ where
the metric function $f(r)$ develops a second -extremal- horizon for  some radius $\tilde r_h$. The radius $\tilde r_h$ and $Q_c$ depend only weakly on $P$~:
quantitatively, we found $ 1 < \tilde r_h / r_h \sim 3.5$ and $Q_c \sim 5.5$.
In this limit, the HBHs then approach a  configurations consisting of two different
regions~: (i) the scalar hair is concentrated in the spherical shell $r_h < r < \tilde r_h$
, the electric field is practically vanishing
and the metric function $a(r)$ is practically constant;
 (ii) the outside of the shell where the scalar hair vanishes and the solution is an extremal (dyonic) Reissner-Nordstrom black hole with horizon $\tilde r_h$.
Such a configuration is illustrated in the left panel of  Fig. \ref{profile_p_05} for $P=0.5, Q_c=5.6640$.
The results suggest that the fields conspire to minimize the Horndeski interaction term  (see the $\alpha$-term in Eq. (\ref{lag_eff}))~: in the inside region the scalar field is large but $a,a'$ and $V'$
get very small; in the outside region $\phi, \phi'$ are very small. Moreover, at the critical solutions of this kind, $\phi' (r)$ diverges as
$r \rightarrow \tilde r_h$ which is consistent with the vanishingly small $f'(\tilde r_h)$ - typical for an ERN. They are both related by the same function ${\cal B}$ as seen in Eqs. (\ref{cond_regular}) and (\ref{expression for f'}). The horizon at $r=r_h$ is still regular as seen from the regularity of the Ricci scalar there.
\item For fixed $Q$ ($0 < Q \lesssim 5.5$),
the increase of $\phi(r_h)$ results for the parameters $Q,P, \phi(r_h)$
to approach the hypersurface ${\cal B}=0$ (see Eqs. (\ref{cond_regular}) and (\ref{expression for B})), which in turn gives rise to the behavior illustrated in the right panel of Fig. \ref{profile_p_05} for $Q=3, P_c=0.7847$. Concretely~: (i)
the scalar field $\phi$ is peaked at the horizon $r_h$ and decreases rapidly for $r > r_h$, (ii)
the parameters $|\phi'(r_h)|$ and $a(r_h)$  diverge rapidly together with $f'(r_h)\rightarrow 0$ - again the ${\cal B}$-function, and  (iii) there are no HBH solutions
beyond a maximal value of $\phi(r_h)$ corresponding to a maximal value $P=P_c$.
This property holds for higher magnetic charge $P \gtrsim 0.76$ and appears along the generally horizontal branch of the blue boundary in Fig. \ref{domain_2}. It suggests that, for these relative values of $Q,P$, the magnetic field wins over the
electric field, enhancing a concentration of the scalar matter nearby the horizon  and preventing the formation of a second horizon. Here too we find a regular Ricci scalar at $r=r_h$, so the horizon stays regular also in this case. Finally, we note that the small domain of small $Q$ and large $P$ between the blue and the red lines exhibits the same behavior.
\item  Moving with the critical solutions along the upper boundary (i.e. increasing electric charge $Q$), we see that the minimal values of $a(r)$ and $f(r)$ become deeper and the profiles become more and more similar to those of the ``vertical'' branch of the boundary. The two types merge at  the tipping point ($Q=6.12$, $P=0.76$).
\end{itemize}

\begin{figure}[h!!!]
\begin{center}
{\includegraphics[width=8cm, angle=-00]{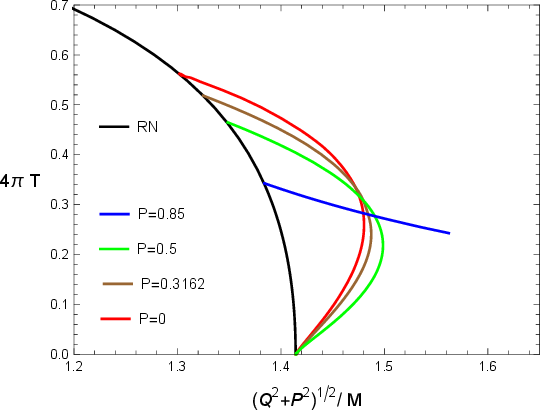}}\hspace{5mm}
{\includegraphics[width=7.8cm, angle=-00]{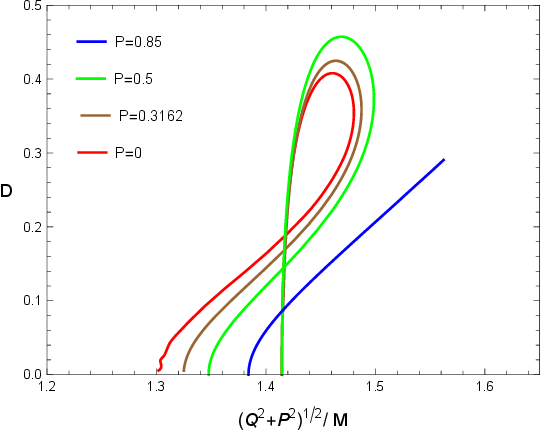}}
\caption{{\small The effect of varying the ratio $\sqrt{Q^2+P^2}/M$ on the HBH temperature $T$ (Left) and on the scalar charge $D$ (Right). The black line in the left panel corresponds to the RN hairless solutions. $\alpha=1$.
\label{DandTvsChargeToMass} }}
\end{center}
\end{figure}

Additional insight into the pattern of the solutions can be obtained by studying the behavior of the scalar charge and the BH temperature as a function of the generalized charge to mass ratio $\varpi=\sqrt{Q^2+P^2}/M$. This behavior is presented in Fig. \ref{DandTvsChargeToMass} and it is easy to observe that all scalarized BH branches start on the RN curve which is just a different parametrization of the red bifurcation curve of Fig. \ref{domain_2}. However, only a subset of the branches end on the RN curve. This is of course another manifestation of  the existence of two different kinds of solutions. The first kind is of the electric type which contains the purely electric HBHs of Ref. \cite{Brihaye:2020oxh} and has the ERN solution as a limiting point with $\varpi= \sqrt{2}$. This kind of solutions start with $D=0$ and return to $D=0$ as an end point. Most solutions between these end points are overcharged, i.e. $\varpi >\sqrt{2}$. The solutions of the second type are dominated by the magnetic field and they are represented by the blue line in Fig. \ref{DandTvsChargeToMass} which corresponds to a family of solutions which terminate in a limiting solution with finite (non-zero) scalar charge and finite temperature. The solutions of this branch can be overcharged to a higher degree. \\

\noindent  {\bf Solutions with $\bf{\alpha}<\textbf{0}$}\\
We will close this presentation by giving some attention to the case of $\alpha<0$. As a general statement, we can say that, in the background of a RN BH,
scalar clouds can form
if the potential $W(r)$ is negative in some region. Looking at the horizon we find
(remember we set $r_h=1$)~:
\be
         W(r_h) = -2\alpha(P^4 + P^2 Q^2 + Q^4 - P^2).
\ee
Therefore, for $\alpha < 0$, HBH are expected to exist in the region whose boundary is the line(s)
\be
Q^2=\frac{1}{2}\bigg(\sqrt{4P^2-3P^4}-P^2\bigg)
\ee
which is bounded by the rectangle $0 < P < 1$ and $0 < Q < 1/\sqrt{3}$ if we assume positive charges.
This is indeed what is found, illustrated by Fig. 2. More precisely, setting $Q=0$ we find that
scalar clouds exist for $ 3.4 < -\alpha < \infty$  (the limit  $-\alpha \to \infty$
corresponding to $P \to 0$).

\begin{figure}[h!!!]
\begin{center}
{\includegraphics[width=10cm, angle=-00]{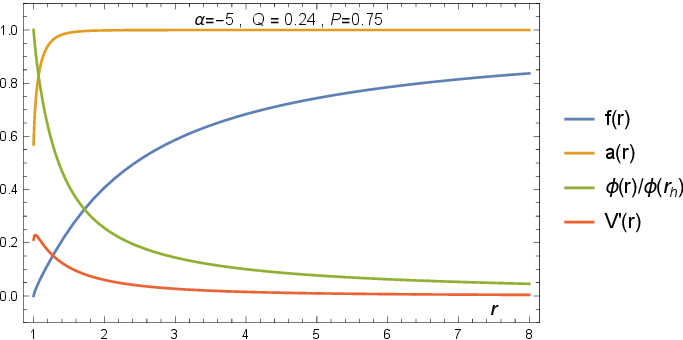}}
\caption{{\small the profiles of a generic HBH with $\alpha<0$. $\alpha=-5$, $P = 0.75, Q= 0.24$.
}}\label{NegativeAlphaProf}
\end{center}
\end{figure}

 In order to confirm the existence of HBH with $\alpha < 0$,
we constructed  families of solutions corresponding to a few generic values of $\alpha, P, Q$
by increasing progressively the parameter $\phi(r_h)$.
For all cases it was found that the HBH solutions approach a
configuration with $a(r_h) = 0$ for some  maximal value of $\phi(r_h)$.
This suggests that the limiting solution is singular at the horizon since, indeed
the Ricci scalar behaves according to
\be
R(r) =  \bigg[-3 \frac{a' f'}{a} + \frac{2}{r_h^2}(1 - 2 r_h f' + r_h^2 f'') \bigg]_{r=r_h}  + o(r-r_h) .
\ee
For example, the HBH corresponding to $\alpha=-5, P = 0.7$  has  $Q \approx 0.2334$;  a HBH family could be constructed  up to $\phi(r_h) \approx 0.235$  corresponding to $Q \sim 0.205$.
The profiles of a generic HBH corresponding to $\alpha=-5$, $P = 0.75, Q= 0.24$ is presented
on Fig. \ref{NegativeAlphaProf} .

\section{Conclusion }
\setcounter{equation}{0}
The work reported by this article is concentrated about BHs with scalar hair which appear naturally in the vector-tensor Horndeski theory coupled (non-minimally) to a real massless scalar field. After analyzing previously \cite{Brihaye:2020oxh} the purely electric system, we study here magnetic and dyonic scalarized BHs of the same system. We find that the addition of magnetic charge changes significantly the properties of these solutions, as might be expected from the fact that the Horndeski term breaks the electromagnetic duality symmetry. The general pattern of scalarization stays the same, namely this system allows RN BHs with vanishing scalar field, but on a certain surface of parameter space  ${\cal S}(\alpha, P, Q) = 0$ these solutions become unstable and scalarized BHs appear. We found that similar to the purely electric scalarized BHs, the analogous purely magnetic scalarized BHs also exist for a finite interval of magnetic field, but the limiting solution with the maximal magnetic field is very different from the electric one which is a hairless extremal RN NH outside a degenerate horizon. In the magnetic case, the RN behavior is achieved only asymptotically and the scalar hair of the critical solution is still observable. Dyonic scalarized BHs exist in a certain domain of parameter space which is bounded by a critical surface
on which the nature of the limiting solutions is determined by the proximity to the $Q=0$ or $P=0$ lines.





\begin{thebibliography}{99}
\bibitem{Horndeski:1974wa}
  G.~W.~Horndeski,
  Int.\ J.\ Theor.\ Phys.\  {\bf 10} (1974) 363.
\bibitem{Horndeski:1976gi}
  G.~W.~Horndeski,
  J.\ Math.\ Phys.\  {\bf 17} (1976) 1980.
\bibitem{muller} F. Muller-Hoissen and R. Sippel,
Class. Quantum Gravity, {\bf 5} (1988) 1473.
\bibitem{Verbin:2020fzk}
  Y.~Verbin,
   ``Magnetic Black Holes in the Vector-Tensor Horndeski Theory,''
  arXiv:2011.02515 [gr-qc].
\bibitem{no_hole_old}
  J.~D.~Bekenstein,
  Phys. Rev. Lett.   {\bf 28} (1972) 452; C. Teitelboim, Lett. Nuovo Cim. {\bf 3S2} (1972) 397.
\bibitem{no_hole_new}
  J.~D.~Bekenstein,
Phys. Rev.  {\bf D 51} (1995)   R6608.
\bibitem{Herdeiro:2014goa}
  C.~A.~R.~Herdeiro and E.~Radu,
  Phys.\ Rev.\ Lett.\  {\bf 112} (2014) 221101.
\bibitem{Sotiriou:2013qea}
  T.~P.~Sotiriou and S.~Y.~Zhou,
  Phys.\ Rev.\ Lett.\  {\bf 112} (2014) 251102.
\bibitem{Sotiriou:2014pfa}
  T.~P.~Sotiriou and S.~Y.~Zhou,
  Phys.\ Rev.\ D {\bf 90} (2014) 124063.
\bibitem{Herdeiro:2015waa}
  C.~A.~R.~Herdeiro and E.~Radu,
  Int.\ J.\ Mod.\ Phys.\ D {\bf 24} (2015)   1542014.
	
\bibitem{Sotiriou:2015pka}
  T.~P.~Sotiriou,
  Class.\ Quant.\ Grav.\  {\bf 32} (2015),  214002.
\bibitem{Volkov:2016ehx}
  M.~S.~Volkov,
  ``Hairy black holes in the XX-th and XXI-st centuries'',
  arXiv:1601.08230 [gr-qc].





\bibitem{Doneva:2017bvd}
  D.~D.~Doneva and S.~S.~Yazadjiev,
  Phys.\ Rev.\ Lett.\  {\bf 120} (2018),  131103.
\bibitem{Silva:2017uqg}
  H.~O.~Silva, J.~Sakstein, L.~Gualtieri, T.~P.~Sotiriou and E.~Berti,
  Phys.\ Rev.\ Lett.\  {\bf 120} (2018),  131104.
\bibitem{Antoniou:2017acq}
  G.~Antoniou, A.~Bakopoulos and P.~Kanti,
  Phys.\ Rev.\ Lett.\  {\bf 120} (2018),  131102.
\bibitem{Antoniou:2017hxj}
G.~Antoniou, A.~Bakopoulos and P.~Kanti,
Phys. Rev. D \textbf{97} 084037 (2018).


\bibitem{Herdeiro:2018wub}
  C.~A.~R.~Herdeiro, E.~Radu, N.~Sanchis-Gual and J.~A.~Font,
  Phys.\ Rev.\ Lett.\  {\bf 121} (2018)   101102	
	
\bibitem{Fernandes:2019rez}
  P.~G.~S.~Fernandes, C.~A.~R.~Herdeiro, A.~M.~Pombo, E.~Radu and N.~Sanchis-Gual,
  Class.\ Quant.\ Grav.\  {\bf 36} (2019)  134002 ;
   Erratum: Class.\ Quant.\ Grav.\  {\bf 37} (2020) 049501.
	

\bibitem{Brihaye:2019kvj}
  Y.~Brihaye and B.~Hartmann,
  Phys.\ Lett.\ B {\bf 792} (2019) 244.

\bibitem{Brihaye:2020oxh}
  Y.~Brihaye and Y.~Verbin,
  Phys.\ Rev.\ D {\bf 102} (2020) 124021.


\bibitem{COLSYS}
 U. Ascher, J. Christiansen, R.~D. Russell,
 Math. Comp. {\bf 33} (1979) 659;
 U. Ascher, J. Christiansen, R.~D. Russell,
 ACM Trans. {\bf 7} (1981) 209.	









\end{thebibliography}
 \end{document}